\documentclass[aps,pre
,showpacs%,unsortedaddress
,twocolumn,twoside
,amsmath,amssymb,amsfonts]{revtex4-1}

\usepackage{txfonts}
\DeclareSymbolFont{largesymbols}{OMX}{cmex}{m}{n}   % ints strange, CM fits fine
\DeclareSymbolFont{letters}{OML}{ztmcm}{m}{it}   %mathptmx' letters (spacing)

\usepackage{graphicx}

\usepackage{microtype}

\DeclareMathOperator{\e}{e}
\newcommand{\eps}{\varepsilon}
\newcommand{\dd}{\mathrm{d}}
\newcommand{\del}{\partial}
\newcommand{\abs}[1]{\left\lvert{#1}\right\rvert}
\renewcommand{\vec}[1]{\mathbf{#1}}
\newcommand{\name}[1]{#1}%{\textsc{#1}}

\def\clap#1{\hbox to 0pt{\hss#1\hss}}
\def\mathclap{\mathpalette\mathclapinternal}
\def\mathclapinternal#1#2{\clap{$\mathsurround=0pt#1{#2}$}}

\DeclareMathAlphabet{\mathcal}{OMS}{cmsy}{m}{n}
\newcommand{\MTE}{\tau}
\newcommand{\PB}{\mathcal{P}}
\newcommand{\EP}{\PB_{0,0}}
\newcommand{\slowT}{T'}
\newcommand{\fpf}{F_\varnothing}
\newcommand{\fpm}{F_\mathrm{M}}

\usepackage{url}   % break long doi or url lines
\usepackage[breaklinks,colorlinks]{hyperref}
\usepackage{breakurl}
\begin{document}

\title{Switching between phenotypes and population extinction}
\author{Ingo Lohmar}\email{lohmar@phys.huji.ac.il}
\author{Baruch Meerson}\email{meerson@cc.huji.ac.il}
\affiliation{Racah Institute of Physics,
  the Hebrew University of Jerusalem,
  Jerusalem 91904, Israel}

\date{\today}

\begin{abstract}
  Many types of bacteria can survive under stress by switching
  stochastically between two different phenotypes: the ``normals'' who
  multiply fast, but are vulnerable to stress, and the ``persisters''
  who hardly multiply, but are resilient to stress.  Previous
  theoretical studies of such bacterial populations have focused on the
  \emph{fitness}: the asymptotic rate of unbounded growth of the
  population.  Yet for an isolated population of established (and not
  very large) size, a more relevant measure may be the population
  \emph{extinction risk} due to the interplay of adverse extrinsic
  variations and intrinsic noise of birth, death and switching
  processes.  Applying a WKB approximation to the pertinent master
  equation of such a two-population system, we quantify the extinction
  risk, and find the most likely path to extinction under both favorable
  and adverse conditions.  Analytical results are obtained both in the
  biologically relevant regime when the switching is rare compared with
  the birth and death processes, and in the opposite regime of frequent
  switching.  We show that rare switches are most beneficial in reducing
  the extinction risk.
\end{abstract}


\maketitle

\section{Introduction}
\label{sec:intro}

Understanding and quantifying the persistence of bacterial populations
is of major importance for the efficient treatment of diseases.
While bacterial persistence was uncovered more than 65 years
ago~\citep{bigger44-treat-staph-infec-penic}, conclusive evidence for
the underlying mechanism was only obtained during the last decade from
laboratory experiments at the single-cell level.
It has been established that an isogenetic population under identical
conditions can still exhibit two different phenotypes.  They are clearly
distinguished by different rates of cell division: ``normals'' multiply
fast, ``persisters'' do it much slower.  For the same reason, however,
normals are much more susceptible to antibiotic treatment, while
persisters are highly resilient to the antibiotic.  An individual
bacterium can switch stochastically (at a certain rate, often without
sensing its environment) between the two phenotypes~\citep{balaban04}
(type-II persistence).

Systems of two interacting subpopulations, such as normals and
persisters, have been studied in different contexts in theoretical
biology~\citep{lachmann96, menu00-bet-hedgin-diapaus-strat-stoch-envir,
  thattai04, wolf05-diver-times-adver}.  Deterministic models of
exponential (unbounded) growth were mostly employed, and analysis
focused on the \emph{fitness}---the time-averaged net growth rate---of
the total population, see, e.g., Refs.~\citep{kussell05a, kussell05b,
  gander07, acar08,
  visco10-switc-growt-microb-popul-catas-respon-envir}.
In favorable conditions, when normal bacteria have a high net growth
rate, frequently switching to persisters is merely a burden, as it
decreases the average net growth.  If the environment changes
(deterministically or randomly) between different states, including some
which represent adverse conditions for the normals, e.g., in the
presence of an antibiotic, the same frequent switching can become
beneficial.  In this case, the persisters uphold a base population size
during such a stress phase, while normals are heavily decimated.
By properly tuning the switching rates between different phenotypic
states, one can optimize the fitness of the total
population~\citep{kussell05b}.  For two phenotypes and two environments,
the average time spent as a certain phenotype should be equal to the
average duration of the environment in which this phenotype is the
fittest one.
In more complicated models (including phenotype-specific response and
recovery times upon a change of the environment) one still finds that,
comparing two (genetic) species, the one with switching rates better
tuned (in the above sense) outperforms the other
fitness-wise~\citep{acar08}.

These are important insights into the role that persisters play in a
growing population.  However, the underlying assumption of exponential
growth is tailored to the description of competition among different
genotypes trying to establish themselves by outgrowing others.  Here
fitness is instrumental to survive in the competition, and a good
indicator of a specific genotypes' prospects.
While such an unbounded growth can be realized \emph{in vitro}, the
necessary resources and space \emph{in vivo} are limited.  To account
for this fact, one should introduce models with bounded
growth~\citep{murray02-mathem-biolog}.  In a deterministic (mean-field)
description, the population will then typically exhibit a stable fixed
point corresponding to an established population.  In addition,
there will be a fixed point describing an extinct population.
In reality, population dynamics is a stochastic process: an established
population is subject to noise coming from the random character of
births and deaths.  A rare chain of events, where deaths dominate over
births, eventually drives an isolated established population into the
absorbing extinction state.
Thus for an isolated established population, the ultimate goal is
survival in the face of intrinsic, and also possibly environmental,
noise.  We suggest, therefore, a paradigm shift in the analysis of
bacterial phenotype switching by focusing on the population extinction
risk.

With this motivation, we consider a simple two-population system of
normals and persisters, possibly in a time-varying environment mimicking
a phase of catastrophic conditions for the population.
In a constant environment, a proper measure of the extinction risk is
the mean time to extinction (MTE) of the population, see,
e.g., Ref.~\citep{ovaskainen10-stoch-model-popul-extin}.
We show that a higher fraction of persisters \emph{exponentially}
increases the MTE even in this setting.
With a transient catastrophic phase, a more informative measure of
extinction risk is the extinction probability increase (EPI)
because of the
catastrophe~\citep{assaf09-popul-extin-risk-after-catas-event}.  Here a
higher fraction of persisters exponentially reduces the EPI.
Therefore, when viewed from the perspective of population extinction
risk, the presence of persisters is always beneficial, providing an
``insurance policy'' against extinction in small communities.  This
should be compared with persisters being a mere burden, unless in
adverse conditions, when viewed from the perspective of fitness.

The remainder of the paper is organized as follows.  In
Sec.~\ref{sec:model-methods} we set up a simple model that describes the
interacting populations of normals and persisters.  We also introduce,
in the same section, the pertinent master equation and employ a WKB
approximation which reduces the master equation to an effective
\name{Hamilton}ian mechanics.  We formulate the mechanical problem that
needs to be solved and describe a numerical iteration method for dealing
with this problem.  Sec.~\ref{sec:instantons} presents a perturbation
theory, based on time-scale separation, first for favorable conditions,
then including a catastrophic phase.  There we obtain approximate
analytic results for the MTE or the EPI, respectively, and for the most
probable path to extinction, and compare them with our numerical
solutions.
In Sec.~\ref{sec:switch-slow} we contrast the biologically relevant
regime of rare switching with the regime of frequently-switching
bacteria.
We discuss the main findings in Sec.~\ref{sec:discussion}.

\section{Model and Methodology}
\label{sec:model-methods}

We consider a well-mixed two-population system the dynamics of which is
described by a continuous-time \name{Markov} process.  The number of
``normals'' is denoted by $n$, that of ``persisters'' by $m$.  Normals
die at a rate that we set to unity throughout, and they multiply at a
rate $B(1-n/N)$ per individual.
In a stochastic model this corresponds to a finite state space, with a
maximum number $n=N$ of normal individuals.  $N$ can be thought of as a
number of sites each of which can carry at most one individual, or as
food resources necessary to produce offspring.  This dynamics coincides
with that of infected individuals in the SIS model, with fixed total
population size $N$, unit recovery rate of infected, and an infection
rate $B/N$ between infected and susceptible
individuals~\citep{weiss71-asymp-behav-stoch-deter-model-epidem}.

We now introduce a persister population whose individuals do not
multiply or die at all.  The populations are coupled by normal
individuals switching to persisters at a rate $\alpha$, and persisters
switching to normals at a rate $\beta$.  The ratio of these switching
rates is denoted $\Gamma=\alpha/\beta$.
In a mean-field description, the average numbers of
individuals are governed by rate equations
\begin{equation}
  \label{RE}
  \begin{aligned}
    \dot n &= B n (1-n/N) - n - \alpha n + \beta m,\\
    \dot m &= \alpha n - \beta m.
  \end{aligned}
\end{equation}
The rate equations have a trivial fixed point (FP) $F_0$ at $n = m = 0$,
which describes population extinction, and a nontrivial FP $\fpm$ at
$n_\mathrm{M} = N(1-1/B)$, $m_\mathrm{M} = \Gamma n_\mathrm{M}$.  A
viable population therefore needs $B>1$, when $\fpm$ is stable, while
$F_0$ is a saddle point.  At the stable FP $\fpm$, the ratio between the
population sizes of persisters and normals is $\Gamma$.

\subsection{Noise and metastability}
\label{sec:noise}

Even for large population size, intrinsic noise is crucial, as it will
ultimately drive the system, residing in the vicinity of the
deterministically stable FP $\fpm$, toward extinction.  The stochastic
system is described by the master equation for the dynamics of the
probability distribution of population sizes, $\PB_{n,m}(t)$,
\begin{equation}
  \label{ME}
  \begin{split}
    &\quad \frac{\dd \PB_{n,m}}{\dd t} = \hat{H} \PB_{n,m} \\
    &= B(n-1) \left(1-\frac{n-1}{N}\right) \PB_{n-1,m}
    -Bn \left(1-\frac{n}{N}\right) \PB_{n,m} \\
    &\quad+ (n+1) \PB_{n+1,m} - n \PB_{n,m} \\
    &\quad
    +\alpha(n+1) \PB_{n+1,m-1} - \alpha n \PB_{n,m} \\
    &\quad+ \beta(m+1) \PB_{n-1,m+1} - \beta m (1-\delta_{n,N}) \PB_{n,m}
    .
  \end{split}
\end{equation}
Here the \name{Kronecker} delta $\delta_{n,N}$ prevents transition to a
state with $n = N+1$.  Together with the prescription $\PB_{n<0,m} = 0 =
\PB_{n,m<0}$ and $\PB_{n>N,m} = 0$, probability is conserved and limited
to the stripe $(n,m)\in[0,N]\times[0,\infty)$.  The extinction
probability $\EP(t)$ is described by the equation
\begin{equation}\label{P00}
  \frac{\dd \EP}{\dd t} = \PB_{1,0}.
\end{equation}
When higher moments are assumed to factorize, the mean-field
equations~\eqref{RE} are recovered by summation over Eq.~\eqref{ME}.

The stochastic system, as described by Eq.~\eqref{ME}, has an absorbing
extinction state $n=0=m$, corresponding to zero eigenvalue and
eigenstate $\delta_{n,0; \,m,0}$ of the transition matrix $\hat{H}$.
All other eigenvalues are negative, hence all other eigenstates of the
probability distribution decay, and the population goes extinct.
We assume (and verify \emph{a posteriori}) that, in contrast to all
other nonzero eigenvalues, the eigenvalue with smallest nonzero absolute
value is exponentially small in the system size $N$.  This corresponds
to a \emph{metastable} distribution centered around
$\fpm$~\citep{dykman94, herwaarden95-stoch-epidem, elgart04, assaf07,
  kamenev08-extin-infec-diseas, dykman08,
  assaf09-popul-extin-risk-after-catas-event,
  khasin09-extin-rate-fragil-popul-dynam,
  khasin10-time-resol-extin-rates-stoch-popul,
  khasin10-speed-up-diseas-extin-limit-amoun-vaccin,
  assaf10-extin-metas-stoch-popul}.  The shape function of this
distribution, normalized to unity, is called the quasistationary
distribution (QSD); we denote it by $\pi_{n,m}$.  The decay time of the
metastable distribution is $\MTE\gg1$.  An initial distribution,
describing a viable population, first quickly relaxes to the QSD on a
time scale $\sim 1/(B-1)$.  Then the metastable distribution will
``leak'' to zero, as described by the equations $\PB_{n,m}(t) \simeq
\pi_{n,m}\exp(-t/\MTE)$ [for $(n,m)\neq(0,0)$] and $\EP(t) \simeq 1-
\exp(-t/\MTE)$, where $\MTE$ is expected to be exponentially large in
$N$.  Using Eq.~\eqref{ME}, the QSD $\pi_{n,m}$ obeys the equation
\begin{equation}
  \label{timesep}
  \hat H\pi_{n,m} = -\pi_{n,m}/\MTE,
\end{equation}
and with $\MTE$ exponentially large in $N$, the right-hand side can be
approximated by zero.
Having found $\pi_{n,m}$, one obtains $\MTE$ by using Eq.~\eqref{P00}:
$\MTE = 1/\pi_{1,0}$.
One can show (see, e.g., Ref.~\citep{assaf07}) that $\MTE$ is indeed the
mean time to extinction (MTE) when starting from the QSD.
We remind the reader that time is measured throughout this paper in
units of the death rate coefficient of the normal population.

\subsection{WKB approximation}
\label{sec:WKB}

When $N$ is sufficiently large, one can approximately solve
Eq.~\eqref{timesep} by a \name{Wentzel-Kramers-Brillouin} (WKB) eikonal
ansatz~\citep{kubo73-fluct-relax-macrov,
  gang87-station-solut-master-eqs-large,
  peters89-station-distr-popul-size-tribol, dykman94}
\begin{equation}\label{WKB-ansatz}
  \pi_{n,m} = \exp\left[-N S(x,y)\right],
\end{equation}
where $x=n/N$ and $y=m/N$ are assumed to be continuous variables.
Having found $S(x,y)$ in the leading order in $1/N$, the MTE can be
calculated up to a pre-exponential factor:
\begin{equation}
\label{MTES}
  \MTE = 1/\pi_{1,0} \approx \exp[N S(0,0)],
\end{equation}
such that $S(0,0)$ plays the role of an entropic barrier against
extinction.

Plugging Eq.~\eqref{WKB-ansatz} into Eq.~\eqref{timesep} and
\name{Taylor}-expanding $S$ around $(x,y)$ to first order, one obtains,
in the leading order of $1/N$, a zero-energy \name{Hamilton-Jacobi}
equation
\begin{equation}\label{HJ}
    H(x, y, \del S/\del x, \del S/\del y) = 0,
\end{equation}
where
\begin{equation}
  \label{hamiltonian}
  \begin{split}
    H(x,y,p_x,p_y)
    &=
    Bx(1-x) \left(\e^{p_x}-1\right)
    +x \left(\e^{-p_x}-1\right) \\
    &\quad
    +\alpha x \left(\e^{-p_x+p_y}-1\right)
    +\beta y \left(\e^{p_x-p_y}-1\right)
  \end{split}
\end{equation}
is the effective \name{Hamilton}ian. The corresponding \name{Hamilton}
equations,
\begin{subequations}\label{hamilton}
  \begin{alignat}{1}
    \label{ham-x}
    \dot x
    &= Bx(1-x) \e^{p_x} - x \e^{-p_x}
    -\alpha x \e^{-p_x+p_y} +\beta y \e^{p_x-p_y}, \\
    \label{ham-y}
    \dot y
    &= \alpha x \e^{-p_x+p_y} - \beta y \e^{p_x-p_y}, \\
    \label{ham-px}
    \dot p_x
    &= - B(1-2x) \left(\e^{p_x}-1\right)
    -\left(\e^{-p_x}-1\right) - \alpha \left(\e^{-p_x+p_y}-1\right), \\
    \label{ham-py}
    \dot p_y
    &= - \beta \left(\e^{p_x-p_y}-1\right),
  \end{alignat}
\end{subequations}
describe trajectories of the system in the four-dimensional phase space
of rescaled population sizes $x$ and $y$ and conjugate momenta $p_x$ and
$p_y$.
To determine $S(x,y)$, one can calculate the mechanical action
accumulated along the proper activation trajectory, or \emph{instanton},
of \name{Hamilton}'s equations of motion and ending in $(x,y)$.

As the \name{Hamilton}ian $H$ does not explicitly depend on time,
$H(x,y,p_x,p_y) = E$ is an integral of motion.  In view of
Eq.~\eqref{HJ}, the energy $E$ must be zero.
One type of motion with $E=0$ occurs in the invariant plane $p_x=p_y=0$
where Eqs.~\eqref{ham-x} and \eqref{ham-y} coincide with the (rescaled)
rate equations~\eqref{RE}.
Overall, there are three zero-energy FPs of the \name{Hamilton}ian flow:
$(0,0,0,0)$, $[1-1/B,\Gamma (1-1/B),0,0]$ and $(0,0,-\ln B,-\ln B)$, all
of them four-dimensional saddles.  The first two originate from the
mean-field FPs, and we will continue referring to them as $F_0$ and
$\fpm$, respectively.  The third FP, which we call $\fpf$, is the
fluctuational extinction point: it appears in a broad class of
stochastic population models exhibiting
extinction~\citep{herwaarden95-stoch-epidem, elgart06,
  kamenev08-extin-infec-diseas, dykman08}.  Note that all the FPs merge
into the origin upon approaching the bifurcation point $B=1$.

As the established population resides around $\fpm$, the instanton must
start at this FP.  Now, as we look for $S(0,0)$, we need to choose
between the fixed points $F_0$ and $\fpf$ as the final destination. It
has been shown that only $\fpf$ can be reached from the region $x$,
$y>0$, $p_x$, $p_y\neq0$~\citep{dykman08,
  khasin10-speed-up-diseas-extin-limit-amoun-vaccin}.  The instanton,
therefore, must be a heteroclinic trajectory which starts at the
metastable FP $\fpm$ at time $-\infty$ and enters the extinction FP
$\fpf$ at time $+\infty$.  Finding the MTE, see Eq.~\eqref{MTES},
demands calculating the action $S = S(0,0)$ along this heteroclinic
trajectory:
\begin{equation}\label{action-forms}
  \begin{split}
    S
    &= \int \dd t\, \left(\vec p\dot{\vec q} - H\right)
    = \int \dd t\, \left(-\dot{\vec p}\vec q - H\right) \\
    &= \int (p_x \,\dd x + p_y \,\dd y - H \,\dd t),
  \end{split}
\end{equation}
where $\vec q=(x,y)$ and $\vec p=(p_x,p_y)$.
In a boundary layer of width $\sim 1/N$ around $x=0$ and $y=0$ the
assumption of large population size $n$, $m\gg1$ breaks down.  However,
for a sufficiently large system size $N$, the contribution of this layer
to the MTE is subleading in the parameter
$1/N$~\citep{kessler07-extin-rates-fluct-induc-metas,
  assaf10-extin-metas-stoch-popul}.

\subsection{Iterative numerical solution}
\label{sec:bnf}

The two-degrees-of-freedom \name{Hamilton}ian \eqref{hamiltonian} has
only one independent integral of motion: the energy.  It is thus
nonintegrable.  Therefore, the instanton can in general be only
obtained numerically.

In earlier work, ``shooting'' algorithms were used to integrate
numerically \name{Hamilton}'s equations of motion for this purpose, see,
e.g., Refs.~\citep{kamenev08-extin-infec-diseas, dykman08,
  assaf09-popul-extin-risk-after-catas-event}.  Below
(Sec.~\ref{sec:cat-num}) we will explain why such an algorithm is not
feasible in our case.  Instead we adapted an iterative algorithm
introduced, in the context of \name{Hamilton}ian field theories, in
Refs.~\citep{chernykh01-large-negat-veloc-gradien-burger-turbul,
  elgart04}.
Let subscripts ``M'' and ``$\varnothing$'' label the initial and the
final FP, respectively.
We fix a sufficiently long calculation time $t_\mathrm{max}$ to traverse
the trajectory; it should not be too long in order to avoid
instabilities in the vicinities of the fixed points.
The starting iteration numerically integrates Eqs.~\eqref{ham-x}
and~\eqref{ham-y} with the momenta fixed at their target values $\vec
p=\vec p_\varnothing$, starting from the initial condition $\vec
q(t=0)=\vec q_\mathrm{M}$ and up to time $t_\mathrm{max}$.
The resulting coordinate curve $\vec q(t)$ is now used to fix the
coordinates in Eqs.~\eqref{ham-px} and~\eqref{ham-py}, leaving a system
of equations for the momenta, which is integrated backwards in
time starting from $\vec p(t=t_\mathrm{max})=\vec p_\varnothing$ down to
$t=0$.
In each following iteration half-step, momenta (coordinates) are fixed
by the time-dependent solution obtained in the previous step, and the
coordinates (momenta) are integrated forward (backward) in time,
starting from the values at the initial (final) FP and up (down) to
$t=t_\mathrm{max}$ ($t=0$).
We found that this scheme rapidly converges to the desired instanton.

To compute the action, we use the expressions in the first line of
Eq.~\eqref{action-forms}.  The difference between these two versions is
an easy measure of the numerical accuracy that has been reached.

This algorithm makes it possible to obtain, with little effort, the most
likely path to extinction and the MTE for a broad class of population
dynamics models when the target FP has a different momentum than the
initial FP (as it happens here).

\section{Instanton Trajectories}
\label{sec:instantons}

\subsection{Close to the bifurcation}
\label{sec:near-bifurc}

To simplify the algebra, we will restrict ourselves to the regime close
to the bifurcation point $B=1$ where all FPs merge, and define the
distance to bifurcation $\delta = B-1 \ll1$.
As can be checked \emph{a posteriori}, $x$, $y/\Gamma$, $\abs{p_x}$,
$\abs{p_y} \sim\delta$ or smaller.  Therefore, exponentials in the
\name{Hamilton}ian~\eqref{hamiltonian} can be \name{Taylor}-expanded.
In addition, we assume that the switching from the normals to persisters
and back is rare: $\alpha$, $\beta\ll\delta\ll1$.
Under these conditions, the \name{Hamilton}ian~\eqref{hamiltonian}
becomes
\begin{equation}\label{hamiltonian-d}
  H(x,y,p_x,p_y)
  \simeq xp_x(p_x-x+\delta)
  -(\alpha x-\beta y)(p_x-p_y).
\end{equation}
Here we neglected terms $\sim\delta^4$, and the term $(\alpha x+\beta
y)(p_x-p_y)^2/2 \sim \alpha\delta^3$.  This is
consistent %since $\delta^3 \gg \alpha\delta^3$, and
if $\alpha \delta^2 \gg \delta^4$, that is, $\delta\ll \sqrt{\alpha}$.
The \name{Hamilton} equations read
\begin{subequations}\label{hamilton-d}
  \begin{align}
    \label{ham-d-x}
    \dot x &=x(2p_x-x+\delta)-(\alpha x-\beta y), \\
    \label{ham-d-y}
    \dot y &= \alpha x-\beta y, \\
    \label{ham-d-px}
    \dot p_x &= -p_x(p_x-2x+\delta) +\alpha(p_x-p_y), \\
    \label{ham-d-py}
    \dot p_y &= -\beta(p_x-p_y),
  \end{align}
\end{subequations}
and the zero-energy FPs are $(0,0,0,0)$ (trivial FP, $F_0$),
$(0,0,-\delta,-\delta)$ (extinction FP, $\fpf$), and $(\delta, \Gamma
\delta, 0,0)$ (metastable FP, $\fpm$).

It is helpful to rescale all quantities by putting $x=\delta X$,
$y=\delta Y$, $p_x=\delta P_X$, $p_y=\delta P_Y$, and $t=T/\delta$.
The equations of motion become
\begin{subequations}\label{rescaled-eom}
\begin{align}
  \frac{\dd X}{\dd T}
  &= X(2P_X-X+1)-\eps(\Gamma X-Y), \\
  \frac{\dd Y}{\dd T}
  &= \eps(\Gamma X- Y), \\
  \frac{\dd P_X}{\dd T}
  &= -P_X(P_X-2X+1) +\eps\Gamma(P_X-P_Y), \\
  \frac{\dd P_Y}{\dd T}
  &= -\eps(P_X-P_Y),
\end{align}
\end{subequations}
where $\eps = \beta/\delta$.  These equations are still canonical with
\name{Hamilton}ian
\begin{equation}
h = H/\delta^3 = XP_X(P_X-X+1) - \eps(\Gamma X-Y)(P_X-P_Y).
\end{equation}
The action becomes $S=\delta^2 s$, where
\begin{equation}\label{rescaled-action}
s = \int (P_X \,\dd X+P_Y \,\dd Y -h \,\dd T).
\end{equation}
The rare-switching limit corresponds to $\eps\ll1$, and we will treat it
perturbatively in the following.

\subsection{Solution in a constant favorable environment}
\label{sec:nocat}

The leading-order behavior of $X$ and $P_X$, the \emph{fast} degrees of
freedom, takes place on the unit time scale $T \sim 1$.  The dynamics of
$Y$ and $P_Y$, the \emph{slow} degrees of freedom, however happens on
the long time scale $T \sim 1/\eps\gg1$.
We formally introduce a separate slow time variable $\slowT = \eps T$ to
account for this separation of time scales, and consider perturbative
solutions of the form
\begin{equation}\label{ms-ansatz}
  \begin{aligned}
    X &= X_0(T) + \eps X_1(T, \slowT) +\dots,\\
    P_X &= P_{X0}(T) + \eps P_{X1}(T, \slowT) +\dots,\\
    Y &= Y_0(\slowT) + \eps Y_1(\slowT) +\dots,\\
    P_Y &= P_{Y0}(\slowT) + \eps P_{Y1}(\slowT) +\dots.
  \end{aligned}
\end{equation}
Inserting into the \name{Hamilton} equations~\eqref{rescaled-eom} yields
a system of partial differential equations in each order of
$\eps$.
Note that, in contrast to previous
work~\citep{kamenev08-extin-infec-diseas,dykman08}, here the dynamics of
fast variables (normals) drives the slow variables (persisters).

\begin{figure}
  \centering
  \includegraphics{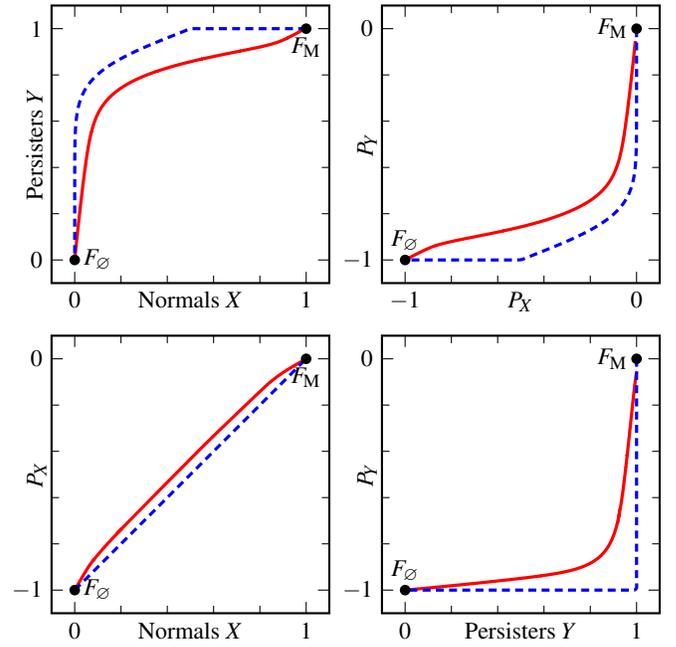}%
  \caption{(Color online) Instanton (constant environment, close to
    bifurcation) for $\Gamma=1$ and $\eps=0.1$ in several projections.
    Theory prediction (dashed blue) and numerical solution (solid
    red).}\label{fig:inst-nocat}
\end{figure}

In the leading order $\sim\eps^0$, only two equations remain, $\dd X_0/
\dd T = X_0(2P_{X0}-X_0+1)$ and $\dd P_{X0}/\dd T =
-P_{X0}(P_{X0}-2X_0+1)$.
This amounts to the \emph{one-dimensional} system of
Ref.~\citep{elgart06, assaf09-popul-extin-risk-after-catas-event} close
to the bifurcation.
The solution must satisfy the energy constraint $h_{X0} =
X_0P_{X0}(P_{X0}-X_0+1) = 0$, hence $P_{X0}=X_0-1$:
the projection of the instanton to the $X$-$P_X$ plane is a straight
line between $\fpm$ and $\fpf$ (cf.\ Fig.~\ref{fig:inst-nocat}), and
this part contributes an action $s_{X0} =
1/2$~\citep{kamenev08-extin-infec-diseas, dykman08,
  assaf09-popul-extin-risk-after-catas-event}.
The solutions for $X_0$ and $P_{X0}$ are
\begin{equation}\label{X-PX-0-d}
  X_0(T) = \frac{1} {1 + \e^{T}},
  \quad
  P_{X0}(T) = \frac{-1} {\e^{-T} + 1},
\end{equation}
where we have arbitrarily fixed the position of the instanton along the
time axis.

The slow persister variables appear in the order $\sim\eps^1$,
\begin{equation}
  \label{Y-PY-eps1}
\begin{aligned}
  \frac{\dd Y_0}{\dd \slowT} +Y_0(\slowT)
  &= \Gamma X_0(T),\\
  \frac{\dd P_{Y0}}{\dd \slowT} -P_{Y0}(\slowT)
  &= -P_{X0}(T).
\end{aligned}
\end{equation}
On the slow time scale of the left-hand sides, the driving terms
$X_0(T)$ and $P_{X0}(T)$ change with time only in the narrow region
$\abs{\slowT}\sim\eps\ll1$; for earlier and later times they are almost
constant.  Therefore, on the slow time scale they can be described as
step functions $X_0 = \theta(-\slowT)$ and $P_{X0} = -\theta(\slowT)$.
We thus solve $\dd Y_0 /\dd \slowT +Y_0(\slowT) = \Gamma
\theta(-\slowT)$ by matching solutions [with $Y_0(-\infty) = \Gamma$,
$Y_0(+\infty) = 0$] at $\slowT=0$,
\begin{equation}
  Y_0(\slowT) =
  \begin{cases}
    \Gamma & \text{for}\ \slowT\leq0,\\
    \Gamma \e^{-\slowT} & \text{for}\ \slowT\geq0.
  \end{cases}
\end{equation}
Similarly, we have $\dd P_{Y0} /\dd \slowT -P_{Y0}(\slowT) =
\theta(\slowT)$ [with $P_{Y0}(-\infty) = 0$, $P_{Y0}(+\infty) = -1$],
such that
\begin{equation}
  P_{Y0}(\slowT) =
  \begin{cases}
    -\e^{\slowT} & \text{for}\ \slowT\leq0,\\
    -1 & \text{for}\ \slowT\geq0.
  \end{cases}
\end{equation}
The phase trajectory projection to the $Y$-$P_Y$ plane forms a rectangle
and contributes an area $s_{Y0} = \Gamma$ to the action.
To resolve the small region $\abs{\slowT}\lesssim\eps$, one would need
to include subleading corrections, which would smoothen the
discontinuous derivatives of $Y_0$ and $P_{Y0}$ at $\slowT = 0$, round
off the trajectory, and decrease the action by small terms $\sim\eps$.

The total action in the leading order $\sim\eps^0$ reads
\begin{equation}\label{s-nocat}
  s_0 = \frac12 + \Gamma.
\end{equation}
The MTE of the population becomes, up to a pre-exponent,
\begin{equation}\label{MTE-nocat}
  \MTE \simeq \exp\left(N\delta^2 s_0\right)
  = \exp\left[N\delta^2 \left(\frac12 + \Gamma\right) \right].
\end{equation}
In comparison, without persisters the MTE is $\simeq \exp\left(N\delta^2
  s_{X0}\right) = \exp\left(N\delta^2/2\right)$, so the persisters cause
an exponential increase of the MTE of the population.  A part of the
exponential increase comes simply from an increased metastable
population size: persisters do not compete with normals, so there is no
``cost'' of increasing their population (via $\Gamma$), only a benefit
against extinction.  Therefore, let us compare the MTE~\eqref{MTE-nocat}
with the MTE $\MTE^\mathrm{1d}$ of a single-population system of
normals, compensated by $N\to N(1+\Gamma)$.  Both systems then have the
same carrying capacity $K = N\delta(1+\Gamma)$.  The ratio of the MTEs
is
\begin{equation}\label{MTE-ratio-nocat}
  \frac{\MTE}{\MTE^\mathrm{1d}}
  = \exp\left[\frac{K\delta\Gamma}{2(1+\Gamma)}\right],
\end{equation}
still exponentially large at $K\delta\gg 1$ and not too small $\Gamma$.
Notable is the effect of increasing the persister fraction
$\Gamma/(1+\Gamma)$ which saturates at large $\Gamma$.
Equation~\eqref{MTE-ratio-nocat} does not suggest any optimal value of
$\Gamma$ but the largest possible one; we will discuss the relation to
other results and the biological context in Sec.~\ref{sec:discussion}.

Interestingly, persisters contribute an action which does not depend on
the absolute switching rates $\alpha$ and $\beta$, see
Eq.~\eqref{s-nocat}.
It may be surprising that an arbitrarily small but finite perturbation
$\eps>0$ yields an exponential change in the MTE with respect to
$\eps=0$.  This is yet another instance of extinction rate
\emph{fragility}~\citep{khasin09-extin-rate-fragil-popul-dynam}.  As in
other ``fragile'' population systems, the explanation to this
counter-intuitive effect comes from a time-resolved
picture~\citep{khasin10-time-resol-extin-rates-stoch-popul}.  The
effective extinction rate is time-dependent.  At relatively small times
$1\ll T\ll 1/\eps$, the extinction rate is the same as if the persisters
were absent ($\eps=0$).  At longer times $T\gtrsim 1/\eps$, the
extinction rate crosses over to its asymptotic value which determines
the
MTE~\eqref{MTE-nocat}~\citep{khasin10-time-resol-extin-rates-stoch-popul}.

In deriving Eq.~\eqref{MTE-nocat}, we assumed closeness to the
bifurcation and rare switching, i.e., $\alpha$, $\beta \ll \delta \ll
1$, or equivalently $\eps$, $\eps\Gamma$, and $\delta \ll 1$; in
particular, implying the upper bound $\Gamma\ll 1/\eps$.
To obtain the approximate \name{Hamilton}ian~\eqref{hamiltonian-d}, we
also had to demand $\alpha\gg\delta^2$ ($\eps\Gamma\gg\delta$); with
hindsight this can be lifted: solving the (effectively one-dimensional)
fast subsystem only employs $\delta\ll1$, while the
ansatz~\eqref{ms-ansatz} only relies on time-scale separation
$\eps\ll1$.  As the small parameters $\delta$ and $\eps$ describe
unrelated mechanisms, the analytical results do not depend (to the given
order) on $\eps\Gamma\gg\delta$.
The WKB approximation is valid, and the resulting MTE $\MTE\gg1$ is
exponentially large, if $N\delta^2 (1/2+\Gamma) \gg 1$.  For that a
minimum system size $N\gg \delta^{-2}$ is sufficient, when $N^{-1/2} \ll
\delta \ll1$ (QSD width much smaller than the distance between initial
and target FPs).

\begin{figure}
  \centering
  \includegraphics{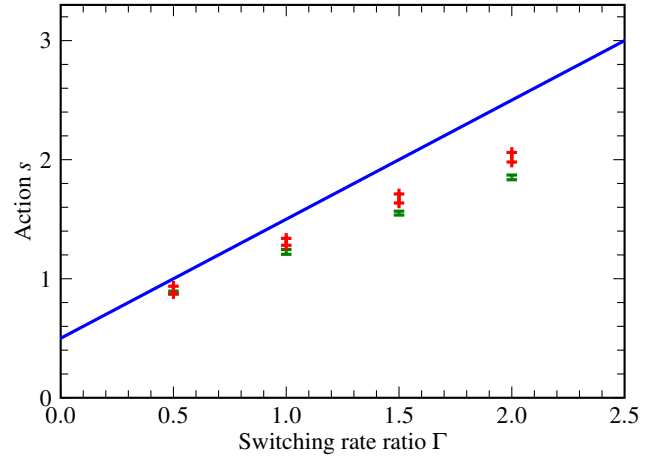}%
  \caption{(Color online) Action $s$ of Eq.~\eqref{rescaled-action}
    versus the ratio of switching rates $\Gamma$,
    analytical~\eqref{s-nocat} (solid blue line) and numerical result
    (green marks $\eps=0.2$, red pluses $\eps=0.1$).  The error bars
    were obtained by using the original action expression and its
    integrated-by-parts counterpart [see
    Eq.~\eqref{action-forms}].}\label{fig:action-nocat}
\end{figure}

Figure~\ref{fig:inst-nocat} compares the instanton found analytically with
the numerical solution (see Sec.~\ref{sec:bnf}) of
Eqs.~\eqref{rescaled-eom} for a moderately small $\eps=0.1$.  Agreement
is reasonably good, and we checked that it improves, in all projection
planes, with decreasing $\eps$.
Figure~\ref{fig:action-nocat} shows that the numerically obtained action
approaches the theoretical value~\eqref{s-nocat} as $\eps\to0$.
The deviation also decreases as $\Gamma$ goes down, as expected.

\subsection{Effect of a catastrophe}
\label{sec:cat-num}

What is the effect of a ``catastrophe'', i.e., temporary adverse
conditions, on the population extinction risk?  For a single population,
this question was addressed in
Ref.~\citep{assaf09-popul-extin-risk-after-catas-event}.  Here we find
that the presence of a persister subpopulation dramatically reduces the
extinction probability increase (EPI) caused by the same type of
catastrophe.

As in Ref.~\citep{assaf09-popul-extin-risk-after-catas-event}, we will
model a catastrophe by setting $B=0$ during a certain period of time
$t_\mathrm c$.  This may mimic the effect of a drug that inhibits cell
multiplication.
The system history then differs from the one described in
Sec.~\ref{sec:noise}.  For early times, after relaxation of the system
to the QSD, the extinction probability still increases with time nearly
linearly as $\EP(t) \simeq 1- \exp(-t/\MTE) \simeq t/\MTE$, where $\MTE$
is the MTE of the system without a catastrophe.  At a time $t_0 \ll
\MTE$, when $\EP=\EP^\mathrm{pre}$, the catastrophe starts, acting for a
duration $t_\mathrm{c} \ll \MTE$.  Compared with $\MTE$, this is a short
transient, which may however considerably increase the extinction
probability to the value $\EP^\mathrm{post}$.  Afterwards, the system is
again described by the (downscaled) QSD and continues to decay, while
the extinction probability increases as $\EP(t) \simeq 1-
(1-\EP^\mathrm{post}) \exp[(t_0+t_\mathrm{c}-t)/\MTE]$.
In this setting, the MTE is too crude a measure of the effect of the
catastrophe: it is dominated by realizations surviving the catastrophe,
resulting in nearly the unperturbed MTE $\MTE$.  Instead, we measure the
influence of the catastrophe by the EPI $\Delta\EP = \EP^\mathrm{post} -
\EP^\mathrm{pre}$.
Up to a pre-exponential factor it is given by
\begin{equation}\label{EPIS}
  \Delta \EP \simeq \e^{-N S_{\mathrm{c}}},
\end{equation}
where $S_{\mathrm{c}}$ is the mechanical action accumulated along the
instanton~\citep{assaf09-popul-extin-risk-after-catas-event}, see
Eq.~\eqref{action-forms}.  While it describes a very different quantity,
one gets, in the leading order, $\Delta \EP$ from the action exactly as
one gets $1/\MTE$ in a constant environment, cf.\ Eq.~\eqref{MTES}.  For
Eq.~\eqref{EPIS} to be valid, in addition to $N S_{\mathrm{c}} \gg 1$
one has to demand that the change of the exponent with respect to
the constant-environment case is large, $N (S - S_{\mathrm{c}}) \gg
1$~\citep{assaf09-popul-extin-risk-after-catas-event}.

The instanton itself is obtained analogously to the case of
time-independent transition rates described in Sec.~\ref{sec:WKB}.
The \name{Hamilton}ian now explicitly depends on time:
Before and after the catastrophe, the system is still described by the
\name{Hamilton}ian~\eqref{hamiltonian}.
During the catastrophe, the effective \name{Hamilton}ian becomes
\begin{equation}\label{hamiltonian-cat}
  H_\mathrm{c}
  = x \left(\e^{-p_x}-1\right)
  + \alpha x \left(\e^{-p_x+p_y}-1\right)
  + \beta y \left(\e^{p_x-p_y}-1\right).
\end{equation}
The instanton trajectory now consists of three connected segments: the
precatastrophe segment starts at the metastable FP $\fpm$ and is
determined by the \name{Hamilton}ian~\eqref{hamiltonian}; the
catastrophe segment is described by Eq.~\eqref{hamiltonian-cat}; the
postcatastrophe segment leads to the extinction FP $\fpf$, again
governed by Eq.~\eqref{hamiltonian}.
We assume that, after the catastrophe ends, there is still a relatively
large population left (with exponentially long MTE).
Neither $H$ nor $H_\mathrm{c}$ depend on time explicitly, therefore on
each segment, energy is conserved: before and after the catastrophe, $H
= E = 0$, and during the catastrophe $H_\mathrm{c} = E_\mathrm{c} \neq
0$.
Furthermore, the phase space points matching the segments are fixed by
the catastrophe duration $t_\mathrm c$.  In turn, this fixes the energy
$E_\mathrm c$.

Again, we consider the system close to the bifurcation, $\delta \ll 1$,
and assume rare switching, $\alpha$, $\beta \ll \delta \ll 1$, such that
before and after the catastrophe we have the
\name{Hamilton}ian~\eqref{hamiltonian-d}.  We expect (and check \emph{a
  posteriori}) that $x$, $y/\Gamma$, $\abs{p_x}$, $\abs{p_y} \sim\delta$
or smaller.
This leads to
\begin{equation}\label{hamiltonian-cat-d}
  H_\mathrm{c}
  \simeq - x p_x + \frac{x p_x^2}{2} - (\alpha x - \beta y) (p_x - p_y),
\end{equation}
where we have kept the same orders as for Eq.~\eqref{hamiltonian-d}.

Rescaling all quantities by $\delta$ as in Sec.~\ref{sec:near-bifurc},
the \name{Hamilton}ian during the catastrophe becomes
\begin{equation}
  h_\mathrm{c} = \frac{H_\mathrm{c}}{\delta^3}
  = - \frac{X P_X}{\delta} + \frac{X P_X^2}{2}
  - \eps(\Gamma X - Y) (P_X - P_Y),
\end{equation}
with the equations of motion
\begin{subequations}\label{hamilton-cat}
  \begin{alignat}{1}
    \label{ham-X-cat}
    \frac{\dd X}{\dd T}
    &= -\frac{X}{\delta} +X P_X -\eps(\Gamma X- Y),\\
    \label{ham-Y-cat}
    \frac{\dd Y}{\dd T}
    &= \eps(\Gamma X- Y), \\
    \label{ham-PX-cat}
    \frac{\dd P_X}{\dd T}
    &= \frac{P_X}{\delta} -\frac{P_X^2}{2} +\eps\Gamma(P_X - P_Y), \\
    \label{ham-PY-cat}
    \frac{\dd P_Y}{\dd T}
    &= -\eps(P_X - P_Y).
  \end{alignat}
\end{subequations}
The rescaled duration of the catastrophe is denoted $T_{\mathrm{c}} =
\delta t_{\mathrm{c}}$.  The leading terms in $\dd X/\dd T$ and $\dd
P_X/\dd T$ are $\sim 1/\delta\gg1$: during the catastrophe the
population size decays exponentially on the fast time scale.

\begin{figure}
  \centering
  \includegraphics{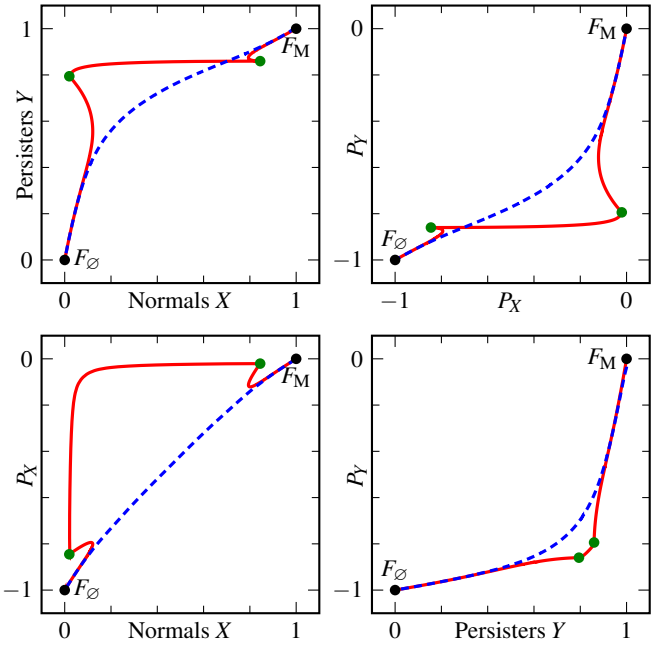}%
  \caption{(Color online) Numerically found instanton for $\Gamma=1$,
    $\eps=0.2$, and $\delta=0.1$, without a catastrophe (dashed blue)
    and with a catastrophe of duration $T_\mathrm c = 0.5$ (solid red).
    Green dots mark both the start and the end of the
    catastrophe.}\label{fig:inst-cat-num}
\end{figure}

To get some insight into the impact of the catastrophe, let us consider
a numerical solution.
To this end, we use the method described in Sec.~\ref{sec:bnf}, where
the equations of motion now change from Eq.~\eqref{rescaled-eom} to
Eq.~\eqref{hamilton-cat} at some time (and back after a duration
$T_{\mathrm{c}}$).
The result is insensitive to this starting time, provided it is
sufficiently far from $t=0$ and $t=t_{\mathrm{max}}$.
Figure~\ref{fig:inst-cat-num} shows several projections of an instanton
with and without the catastrophe phase, for otherwise identical
parameters.
In the top panels, due to time-scale separation the catastrophe segment
is nearly horizontal---$X$ and $P_X$ rapidly decay, persisters are
(indirectly) affected much later.
The bottom panels show that a subpopulation size and its conjugate
momentum do not change simultaneously.  For persisters, first the
momentum builds up, then the population size drops, as in a constant
environment (see Fig.~\ref{fig:inst-nocat}).  For normals, on the other
hand, the situation has changed; the population size now decays earlier
than the momentum, this will be explained in Sec.~\ref{sec:cat-ana}.
The sudden onset and end of the catastrophe is reflected by
nonsmoothness of the instanton (except for the $Y$-$P_Y$ projection).
``Wiggles'' due to nonmonotonic $X$ and $P_X$ immediately precede or
follow the catastrophe segment (we confirmed that these are not
numerical artifacts).
One can see that, after an initial decay of the normal subpopulation
size, it briefly recovers, only to be hit all the harder by the
catastrophe.  Afterwards there is a short recovery period caused by
influx from persisters (cf.\ the $Y$-$P_Y$ projection).

The two-population system with a catastrophe shows a fundamental
difference from the single-population case: the instanton is not only
changed during the catastrophe phase, but the \emph{whole trajectory}
including pre- and postcatastrophe segments is affected.  This can be
understood via the following counting argument.

Imagine we try to match, in a $d$-population system with piecewise
constant \name{Hamilton}ian, the three segments of the instanton.
The $2d$-saddle $\fpm$ affords a $d$-dimensional unstable manifold of
possible end points of the precatastrophe segment.  A useful
parametrization of this point (where the catastrophe segment begins)
consists in $d-1$ ``angles'' describing different trajectories, and a
time-like parameter along the trajectories.
By matching the catastrophe segment of a given duration, the phase space
point at its end is then fixed as well.
At the other end, $\fpf$ affords a $d$-dimensional stable manifold of
possible starting points for the postcatastrophe segment, which can be
parametrized as above.
We thus have $d+d$ parameters at our disposal describing the possible
points at the end of the catastrophe segment and the start of the
postcatastrophe segment.  Since we have to match them in $2d$ phase
space coordinates, this picture does not contradict a unique instanton
(although there still may be more than one solution).

For the single-population case $d=1$, the phase trajectories leaving and
entering the fixed points are unique, and thus cannot be affected
by the catastrophe part in between.  In the generic case $d\geq2$,
however, the pre/postcatastrophe segments may differ from the
no-catastrophe instanton.
In the concrete model studied here, these segments have to differ
simply because of time-scale separation.  During the catastrophe,
normals are rapidly decimated, whereas the persister dynamics follows
much more slowly.  In the $X$-$Y$-projection, the catastrophe segment is
thus less steep than the slope between any two points on the
no-catastrophe instanton.  It is therefore impossible to simply splice
the catastrophe segment into the latter.

This explains why ``shooting'' algorithms are impractical for finding
the catastrophe-related instanton numerically in a multipopulation
system.
For a single population with
catastrophe~\citep{assaf09-popul-extin-risk-after-catas-event}, such an
algorithm can start with a small displacement from the metastable FP
along the \emph{no-catastrophe} instanton, testing different starting
points of the catastrophe segment---this works as the precatastrophe
segment is unchanged.
Likewise, one can parametrize the zero-energy trajectories leaving the
initial FP in the two-population system without catastrophe (see
Sec.~\ref{sec:nocat}) by a shooting angle.
Adding a catastrophe provides an additional freedom (in the form of the
starting point), and the method is no longer practical.

At the same time, Fig.~\ref{fig:inst-cat-num} shows that the instantons
without and with catastrophe practically coincide (in all projections)
for an extended part next to both FPs, before eventually departing from
each other.  This means that the system is extremely sensitive to minute
variations in the angle at which the trajectory leaves (enters) the
initial (final) FP, which only become visible closer to the catastrophe
segment.  We confirmed this behavior in tests of the aforementioned
shooting algorithm (without catastrophe), which, for this reason,
already proves to be rather tedious.

\subsection{Analytic theory with catastrophe}
\label{sec:cat-ana}

We look for an analytic solution analogously to Sec.~\ref{sec:nocat}.
Time-scale separation is still effective: $X$ and $P_X$ show fast
dynamics on the time scale $T\sim1$, or even $T\sim\delta$, see
Eq.~\eqref{hamilton-cat}.  They drive the slow $Y$ and $P_Y$, which
change on a time scale $\slowT=\eps T$.  We denote the catastrophe
duration on this scale by $\slowT_\mathrm{c} = \eps T_\mathrm{c}$.

The leading-order equations $\sim\eps^0$ reduce to the normals-only
system again, and $h_{X,\mathrm c} = -X P_X/\delta + X P_X^2/2$ governs
the dynamics during the catastrophe.  Since $X$, $P_X\sim1\ll1/\delta$,
we neglect the second term, and arrive at the simple catastrophe
\name{Hamilton}ian $h_{X,\mathrm c} \simeq -X P_X/\delta$ used in the
single-population
model~\citep{assaf09-popul-extin-risk-after-catas-event}.  The solution
is an exponential decay (growth) of $X$ ($P_X$) at a rate $1/\delta$ and
for a duration $T_\mathrm c$.  Let $X^+ > X^-$ and $0 > P_X^+ > P_X^-$
denote coordinates and momenta at the start and the end of the
catastrophe, respectively.  Then $X^- = X^+ \exp(-T_\mathrm c/\delta)$
and $P_X^- = P_X^+ \exp(+T_\mathrm c/\delta)$.
The solution for $X$ and $P_X$ before and after the catastrophe is the
same (up to a time shift) as in the constant environment,
Sec.~\ref{sec:nocat}.  This is no contradiction to the arguments of
Sec.~\ref{sec:cat-num}, since the leading approximation is effectively
one-dimensional.  Therefore,
\begin{equation}
  X_0(T) =
  \begin{cases}
    \left(1+\e^{T-T_<}\right)^{-1} & \text{for}\ X_0\geq X^+,\\
    \left(1+\e^{T-T_>}\right)^{-1} & \text{for}\ X_0\leq X^-,
  \end{cases}
\end{equation}
and $P_{X0}(T) = X_0(T)-1$ for both $P_{X0} \geq P_X^+ = X^+-1$ and
$P_{X0} \leq P_X^- = X^--1$.  The quantities $T_<$ and $T_>$ are yet
undetermined.
From the constraints, we get
\begin{equation}
  X^\pm = \frac{1}{1+\e^{\mp T_\mathrm c/\delta}},
  \quad
  P_X^\pm = \frac{-1}{1+\e^{\pm T_\mathrm c/\delta}},
\end{equation}
and the conserved ($X$-part) energy during the catastrophe becomes
$h_{X0,\mathrm{c}} = \cosh^{-2}[T_\mathrm c/(2\delta)] / (4\delta)$.
Fixing the time such that the catastrophe occurs between $T = \pm
T_\mathrm c/2$, we obtain
\begin{equation}\label{X-0-cat}
  X_0(T) =
  \begin{cases}
    \left(1+ \e^{T - T_\mathrm c(1/\delta - 1/2)}\right)^{-1}
    & \text{for}\ T \leq -\frac{T_\mathrm c}2
    ,\\
    % X^+ \e^{-(T+T_\mathrm c/2)/\delta}
    % = X^- \e^{-(T-T_\mathrm c/2)/\delta}
    \frac{\exp(-T/\delta)}{2\cosh[T_\mathrm c/(2\delta)]}
    & \text{for}\ -\frac{T_\mathrm c}2 \leq T \leq +\frac{T_\mathrm c}2,\\
    \left(1+ \e^{T + T_\mathrm c (1/\delta -1/2)}\right)^{-1}
    & \text{for}\ \frac{T_\mathrm c}2 \leq T.\\
  \end{cases}   % frac because of linewidth
\end{equation}
The momentum is $P_{X0} = X_0 -1$ before and after the catastrophe, and
during it decays as
\begin{equation}
  P_{X0}(T) =
  % P_X^+ \e^{(T + T_\mathrm c/2)/\delta}
  % = P_X^- \e^{(T - T_\mathrm c/2)/\delta}
  \frac{-\e^{T/\delta}}{2\cosh[T_\mathrm c/(2\delta)]} = -X_0(-T).
\end{equation}
The action found for this \name{Hamilton}ian and trajectory is
$s_{X0,\mathrm{c}} = \left[1+ \exp(T_\mathrm
  c/\delta)\right]^{-1}$~\citep{assaf09-popul-extin-risk-after-catas-event}.
During the catastrophe, the ``trajectory contribution'' $\int P_{X0}
\,\dd X_0$ and $\int - h_{X0, \mathrm{c}} \,\dd T$ cancel each other.

The slow equations of motion \eqref{ham-Y-cat} and \eqref{ham-PY-cat}
are the same as in the favorable environment of Sec.~\ref{sec:nocat},
hence the slow leading-order equations~\eqref{Y-PY-eps1} (and boundary
conditions) are unchanged.
Again, we only resolve the slow dynamics here.
The driving terms $X_0$ and $P_{X0}$ are different now, since a part of
their movement is replaced by a faster exponential decay (rate
$1/\delta\gg1$) during the catastrophe.  Therefore on the slow $\slowT$
scale one obtains a step function as an even better approximation than
in Sec.~\ref{sec:nocat}.
The only difference between Eqs.~\eqref{X-PX-0-d} and~\eqref{X-0-cat} is
that the driving by $X_0$ ($P_{X0}$) ends (sets in) at the start (end)
of the catastrophe $T=\mp T_\mathrm c/2$ (instead of $T=0$), such that
\(X_0 = \theta( -\slowT_\mathrm c/2 - \slowT )\) and \(P_{X0} = -
\theta(\slowT-\slowT_\mathrm c/2)\).

Since coordinates and momenta remain separate in Eqs.~\eqref{Y-PY-eps1},
the general piecewise solutions for $Y_0$ and $P_{Y0}$ are unchanged,
but now matched at $\slowT = \mp\slowT_\mathrm c/2$:
\begin{equation}
  Y_0(\slowT) =
  \begin{cases}
    \Gamma & \text{for}\ \slowT \leq -\slowT_\mathrm c/2, \\
    \Gamma \e^{-\slowT -\slowT_\mathrm c/2}
    & \text{for}\ -\slowT_\mathrm c/2 \leq \slowT,
  \end{cases}
\end{equation}
and
\begin{equation}
  P_{Y0}(\slowT) =
  \begin{cases}
    -\e^{\slowT -\slowT_\mathrm c/2}
    & \text{for}\ \slowT \leq \slowT_\mathrm c/2, \\
    -1 & \text{for}\ \slowT_\mathrm c/2 \leq \slowT.
  \end{cases}
\end{equation}
The simple geometric picture that the catastrophe merely time-shifts
$Y_0$ and $P_{Y0}$ into opposite directions results in a hyperbola
$Y_0P_{Y0} = -\Gamma \exp(-\slowT_\mathrm c)$ on the corresponding
segment.

Persisters contribute an action
\begin{equation}
  s_{Y0, \mathrm{c}} = \int P_{Y0} \,\dd Y_0 - \int
  h_{Y0, \mathrm{c}} \,\dd T,
\end{equation}
with the switching \name{Hamilton}ian $h_{Y,\mathrm c} = -\eps(\Gamma X
- Y) (P_X - P_Y)$.
The energy during the catastrophe is evaluated on the slow time scale,
such that $X_0 = 0 = P_{X0}$, and
\begin{equation}
  h_{Y0, \mathrm{c}} = - \eps (\Gamma X_0 - Y_0) (P_{X0} - P_{Y0})
  = \eps \Gamma \e^{-\slowT_\mathrm c}.
\end{equation}
The contribution to the action $-h_{Y0, \mathrm{c}} T_\mathrm c = -
\Gamma \slowT_\mathrm c \exp(-\slowT_\mathrm c)$ again cancels the phase
space area under the catastrophe segment, $\int_{\Gamma}^{\Gamma \exp(
  -\slowT_\mathrm c)} P_{Y0} \,\dd Y_0$.  Hence the persister action is
$s_{Y0, \mathrm{c}} = \Gamma \exp(-\slowT_\mathrm c)$, and the total
action becomes
\begin{equation}\label{s-cat}
  s_{0, \mathrm{c}} = \frac{1}{1+\e^{T_\mathrm c/\delta}}
    +\Gamma \e^{-\slowT_\mathrm c}.
\end{equation}
Reinstating the original time scale $t$ by using $\slowT = \eps T =
\eps\delta t = \beta t$ we obtain from Eq.~\eqref{EPIS}
\begin{equation}\label{EPI-ana}
  \Delta \EP \simeq \exp\left[-N\delta^2 \left(\frac{1}{1+\e^{t_\mathrm c}}
    +\Gamma \e^{-\beta t_\mathrm c}\right)\right].
\end{equation}
The system without persisters ($\Gamma=0$) has an EPI $\simeq
\exp\left(-N\delta^2 s_{X0,\mathrm c} \right) = \exp\left[-N\delta^2
  /\left(1+\e^{t_\mathrm c}\right)\right]$.  As for favorable
conditions, we compare with the EPI $\Delta \EP^\mathrm{1d}$ of such a
single-population system of normals, compensated by $N\to N(1+\Gamma)$
to have the same carrying capacity $K=N\delta(1+\Gamma)$:
\begin{equation}\label{EPI-ratio-cat}
  \frac{\Delta \EP}{\Delta \EP^\mathrm{1d}}
  = \exp\left[-\frac{K\delta \Gamma}{1+\Gamma}
    \left(\e^{-\beta t_\mathrm c} - \frac{1}{1+\e^{t_\mathrm c}}\right) \right].
\end{equation}
The system with persisters has exponentially smaller EPI, to which the
initial population size $K$ and the persister fraction
$\Gamma/(1+\Gamma)$ contribute as to the MTE
ratio~\eqref{MTE-ratio-nocat}.  The parenthesized factor quantifies the
fundamental benefit of persisters and generalizes the numerical value
$1/2$ in Eq.~\eqref{MTE-ratio-nocat}: the effect is most pronounced for
catastrophes which are long on the fast scale of normals, but short on
the slow persister time scale, $t_\mathrm c \gg 1 \gg \slowT_\mathrm c =
\beta t_\mathrm{c}$.  Then $\Delta \EP/\Delta \EP^\mathrm{1d} \simeq
\exp\left[-K\delta \Gamma/ \left(1+\Gamma\right) \right]$, i.e., the
ratio is squared with respect to the MTE ratio~\eqref{MTE-ratio-nocat}
in a constant favorable environment: the benefit of persisters is even
more apparent in the face of a catastrophe.
Again the result~\eqref{EPI-ratio-cat} suggests to choose $\Gamma$ as
large as possible, on which we comment in Sec.~\ref{sec:discussion}.

These results are based on $\delta$, $\eps$, $\eps\Gamma\ll1$ (cf.\ the
end of Sec.~\ref{sec:nocat}).
For a short catastrophe $t_\mathrm c\sim 1$ or smaller, the WKB
result~\eqref{EPIS} is valid if the reduction $N \delta^{2}
(s_{0}-s_{0,\mathrm{c}})$ due to the catastrophe is sufficiently large,
yielding the condition $N \gg 4\delta^{-2} / t_{\mathrm{c}}$.
A long catastrophe $\slowT_\mathrm c\sim1$ (or larger) strongly reduces
the action, and the stricter condition is that the remaining action be
large enough.  Considering $\Gamma\sim1$ for simplicity, the persister
action dominates, leading to $N\gg \exp(\slowT_\mathrm c) /
(\delta^2\Gamma)$.

\begin{figure}
  \centering
  \includegraphics{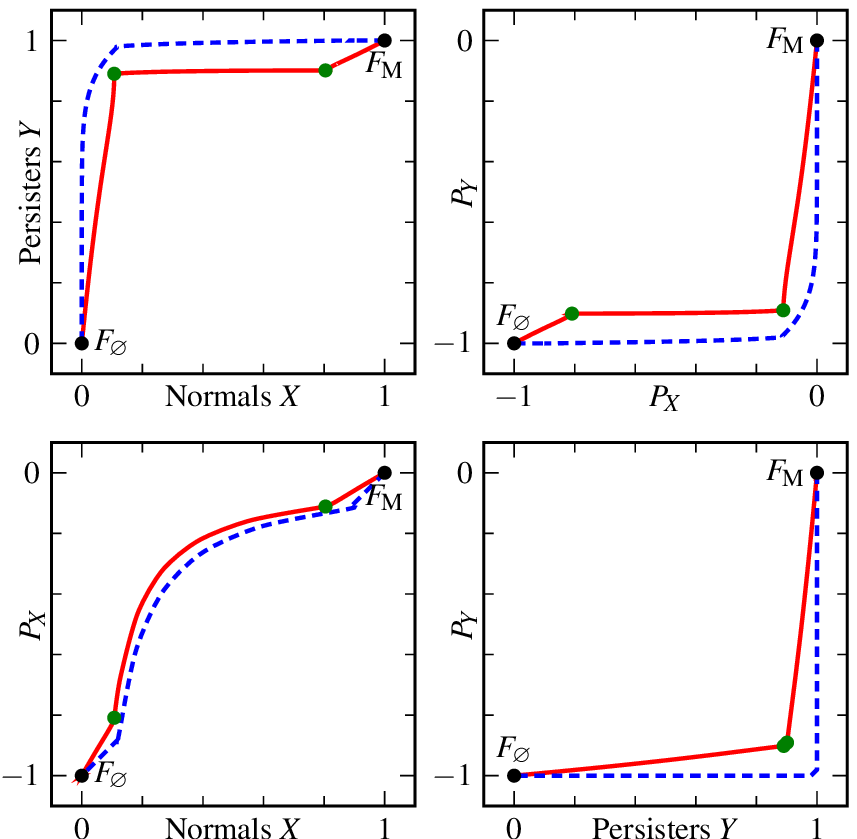}\\[7mm]
  \includegraphics{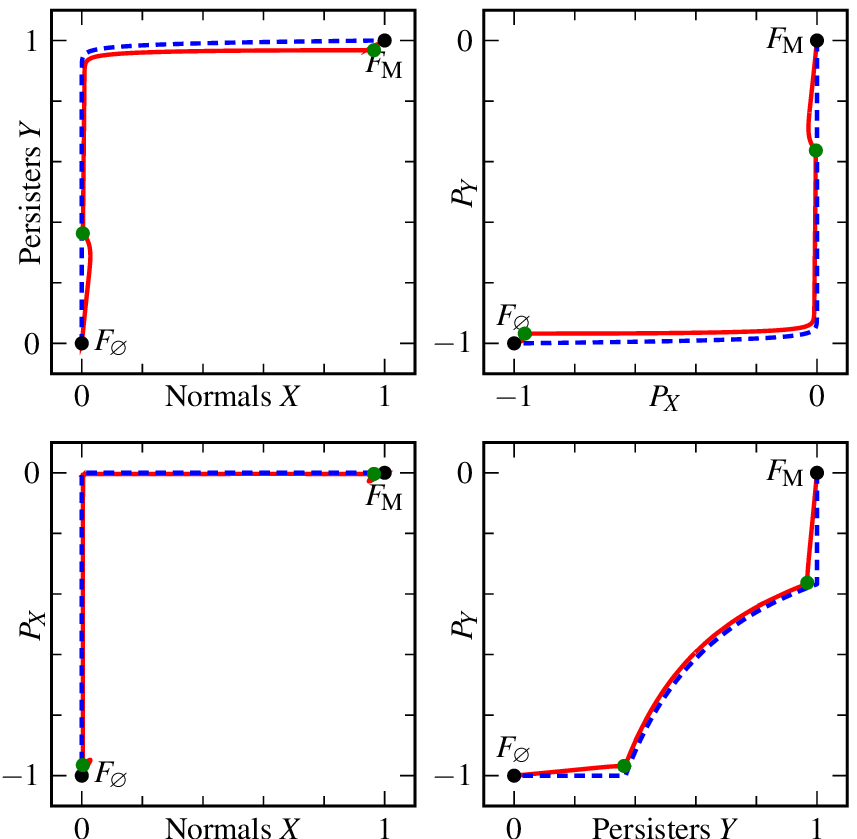}%
  \caption{(Color online) Instanton for $\Gamma=1$, $\eps=0.1$,
    and $\delta=0.1$, with a catastrophe of duration $T_\mathrm c=0.2$
    (top), $10$ (bottom), respectively.  Prediction by theory (dashed
    blue) and numerical solution (solid red).  Green dots mark both the
    start and the end of the catastrophe.}\label{fig:inst-cat-ana}
\end{figure}

\begin{figure}
  \centering
  \includegraphics{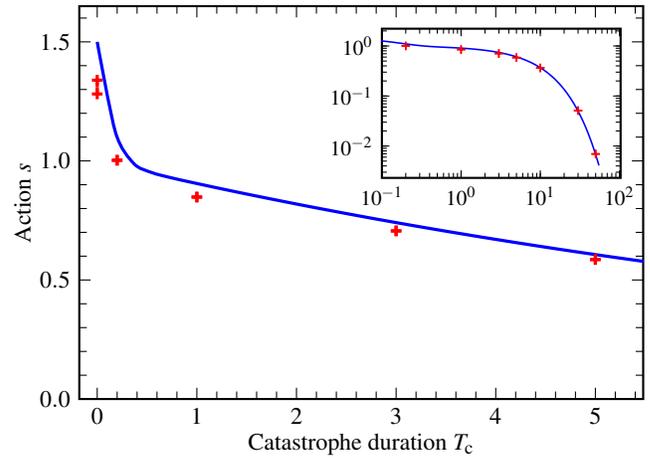}%
  \caption{(Color online) Action $s$ versus catastrophe duration
    $T_\mathrm c$, analytical~\eqref{s-cat} (solid blue line) and
    numerical result (red pluses), for $\Gamma=1$, $\eps=0.1$,
    and $\delta=0.1$.  Error bars span the results obtained using the
    original action expression and its integrated-by-parts
    counterpart.}\label{fig:action-cat}
\end{figure}

The theory path to extinction is shown in Fig.~\ref{fig:inst-cat-ana}
and compared with the numerical solution (see
Secs.~\ref{sec:bnf},~\ref{sec:cat-num}).
For a short catastrophe $T_\mathrm c=0.2$, persisters are mostly
unaffected, while the $X$-$P_X$ projection resembles the one-dimensional
system~\citep{assaf09-popul-extin-risk-after-catas-event}.  Already for
the moderate $T_\mathrm c=1$ (not shown), normals have gone virtually
extinct at the end of the catastrophe, and the population survives
mainly due to the remaining persisters.  With a long catastrophe
$T_\mathrm c=10$, the action contributed by persisters is severely
decreased as well.
Agreement between analytical and numerical solutions is better than in a
constant environment.  Normals go extinct nearly exclusively during the
catastrophe, which completely determines the fast part of the
trajectory, rendering the instanton very simple.  In turn, back-reaction
of persisters becomes less important, and replacing the fast driving
terms by step functions on the slow time scale becomes more accurate.
These are the main approximations of the zeroth-order theory, hence the
predictions improve with increasing catastrophe duration.
We also confirmed that in all projections, the theory becomes more
accurate with decreasing $\eps$.  At the same time, the ``wiggles''
identified in Sec.~\ref{sec:cat-num} become less pronounced.  Both
tendencies go hand in hand, as both are based on reducing back-reaction.
In Fig.~\ref{fig:action-cat}, we compare the action~\eqref{s-cat} with
numerical results.  Even for moderately rare switching ($\eps=0.1$), the
analytical prediction is reasonably accurate, and improving with
increasing catastrophe duration.

We summarize the effect of the catastrophe in the leading order of rare
switching.
Independent of its duration, the strength of the catastrophe is
set by the (normalized) death rate of normals.  Normals decay
exponentially on the very fast scale $t\sim1$, responsible for the major
part of phase space motion (unless $t_\mathrm c\ll 1$).  Persisters are
affected indirectly via switching between the two populations.
For a short catastrophe, $\slowT_\mathrm c \ll 1$, the effect on $Y_0$
and $P_{Y0}$ is negligible: switching hardly occurs during
$\slowT_\mathrm{c}$, and the slow dynamics cannot resolve the difference
in driving.  Therefore only the normal action is reduced, and persisters
are perfectly buffered against the catastrophe.  Note that the time
$t_\mathrm c \ll 1/(\delta\eps)$ can be much longer than the typical
lifetime of an individual normal $\sim 1$.
If the catastrophe is long enough to be seen on the slow scale,
$\slowT_\mathrm c \sim 1$ or larger, switching has an effect.  While
persisters still cannot resolve the accelerated extinction of normals,
they trace the delay between $X_0$ and $P_{X0}$ in the instanton.  On
the slow switching time scale it appears far shorter, forming a buffer
that mitigates the catastrophe.
The structure of the EPI~\eqref{EPI-ana} is thus based on the
separation between the time scale of the catastrophe effect
(strength), and the far longer time scale of persister dynamics.  The
catastrophe affects both populations, but acting on normals, its
duration is measured on the very fast scale $t\sim1$ of the death rate
[action scales $\sim\exp(-t_\mathrm c)$]; acting on persisters it is
measured on the slow scale $\slowT\sim1$ of switching back to normals
[$\sim\exp(-\slowT_\mathrm c)$].  The crossover shows prominently in
Fig.~\ref{fig:action-cat}.

\section{Why Are Switching Rates Small in Nature?}
\label{sec:switch-slow}
So far, we have considered small switching rates between the normal and
persister states, $\eps\ll1$.  The corresponding time-scale separation
was the basis of our qualitative explanation and analytical treatment of
the system's dynamics.
We also numerically examined what happens at $\eps\sim1$ or larger.  We
still consider the system described by the
\name{Hamilton}ians~\eqref{hamiltonian-d} and~\eqref{hamiltonian-cat-d},
respectively, as motivated at the end of Sec.~\ref{sec:nocat}.  The
instanton and the associated action are again obtained as detailed in
Secs.~\ref{sec:bnf},~\ref{sec:cat-num}.

We found that both with and without catastrophe, instanton trajectories
are qualitatively similar to the $\eps\ll1$ case even when $\eps=1$.
Further increasing $\eps$ ``locks'' persisters ever stronger to the
dynamics of normals, see Eqs.~\eqref{rescaled-eom}
and~\eqref{hamilton-cat}.  For very large $\eps$, $P_Y \simeq P_X$ and $Y
\simeq \Gamma X$ with only small deviations.
Moreover, persisters still increase the action compared with a
normals-only system of the same carrying capacity.
We examined the action as a function of varying switching rate $\eps$
and catastrophe duration $t_\mathrm{c}$ ($N$, $\delta$ and $\Gamma$, and
hence the carrying capacity $K$, being fixed).
As expected, for given $\eps$ the action decreases with increasing
catastrophe duration $t_\mathrm{c}$, and this decrease becomes stronger
for larger switching rate $\eps$: the more frequent the switching is,
the less insurance against extinction persisters provide.
For given $t_\mathrm{c}$, the action decreases with increasing switching
rate $\eps$, and this decrease becomes stronger for longer catastrophe
duration: persisters are especially beneficial in the face of a
catastrophe.

For very frequent switching, there is a new time-scale separation which
permits an analytical treatment.
Consider the case $\delta \ll 1 \ll \alpha$, $\beta$, ($\eps$,
$\eps\Gamma \gg 1/\delta$), such that switching is frequent compared
with the normal dynamics even during the catastrophe.
In both the favorable [see Eqs.~\eqref{rescaled-eom}] and catastrophic
[see Eqs.~\eqref{hamilton-cat}] environments, we have
\begin{equation}
  Y = \Gamma X - \frac{1}{\eps} \frac{\dd Y}{\dd T},
  \qquad
  P_Y = P_X + \frac{1}{\eps} \frac{\dd P_Y}{\dd T}.
\end{equation}
For large $\eps$ the second term is a small correction, and we obtain
\begin{equation}\label{fast-persisters}
  Y = \Gamma X - \frac{\Gamma}{\eps} \frac{\dd X}{\dd T} + \dots,
  \qquad
  P_Y = P_X + \frac{1}{\eps} \frac{\dd P_X}{\dd T} + \dots.
\end{equation}
Inserting into the normal equations of motion yields, in the leading
order in $1/\eps$, the normals-only equations, but with a rescaled
time $\tilde T=T/(1+\Gamma)$:
\begin{equation}
  \frac{\dd X}{\dd \tilde T} = X(2P_X-X+1),
  \qquad
  \frac{\dd P_X}{\dd \tilde T} = -P_X(P_X-2X+1)
\end{equation}
in a favorable and
\begin{equation}
  \frac{\dd X}{\dd \tilde T} = -\frac{X}{\delta} +X P_X,
  \qquad
  \frac{\dd P_X}{\dd \tilde T} = \frac{P_X}{\delta} -\frac{P_X^2}{2}
\end{equation}
in a catastrophic environment.
As in Sec.~\ref{sec:cat-ana}, from this we get $X$ as of
Eq.~\eqref{X-0-cat}, only with the substitutions $T_{(\mathrm{c})} \to
T_{(\mathrm{c})}/(1+\Gamma)$, and likewise for the momentum $P_X$ and
the energy $h_{X, \mathrm{c}}$.  $Y$ and $P_Y$ are given by
Eqs.~\eqref{fast-persisters}.

Calculating the action along this instanton, first note that the
switching term in the \name{Hamilton}ian is $h_{Y(,\mathrm{c})} \sim
1/\eps$ at all times.
Second, the corrections in Eq.~\eqref{fast-persisters} do not contribute
to the leading-order action, which becomes
\begin{equation}
  \begin{split}
  s_{\mathrm{c}}
  &\simeq \int\limits_{\mathclap{\text{pre\,/\,post}}}
  (P_X \,\dd X + P_Y \,\dd Y)
  + \int\limits_\text{cat.} (P_X \,\dd X + P_Y \,\dd Y
  - h_{X,\mathrm{c}} \,\dd T) \\
  &\simeq (1+\Gamma) \int\limits_{\mathclap{\text{pre\,/\,post}}} P_X \,\dd X
  + (1+\Gamma) \int\limits_\text{cat.} P_X \,\dd X
  - h_{X,\mathrm{c}} T_\mathrm{c}.
  \end{split}
\end{equation}
The second and third terms cancel; factoring out $(1+\Gamma)$, both
contributions are the same as in the rare-switching case, only with the
above rescaling applied to \emph{all} times.
The first integral is also known from the rare-switching case, where it
coincided with the total action contributed by normals.  Applying
the time rescaling, the action thus becomes
\begin{equation}\label{s-fast-persisters}
  s_{\mathrm{c}} \simeq \frac{1+\Gamma}{1+\e^{T_\mathrm{c}/[\delta(1+\Gamma)]}}.
\end{equation}
We confirmed (for $\eps=100$ and various values of $\Gamma$ and
$t_\mathrm{c}$) that this agrees excellently with the action found
numerically as described at the beginning of this section.
This result is easily interpreted; very frequent switching effectively
``mixes'' the two subpopulations, as they rapidly switch back and
forth.  Compared with a normals-only system, the factor $1+\Gamma$ in
the numerator describes the increased size of the combined population.
A more subtle effect is the reduction, by the same factor $1+\Gamma$, of
the effective duration of the catastrophe.  This reduction accounts for
the lag still gained by switching to the persister state.

In a favorable environment ($t_\mathrm{c} = 0$), persisters switching
frequently do not provide any benefit compared with a normals-only
system of the same carrying capacity.  With a catastrophic phase,
however, we obtain
\begin{equation}
  \frac{\Delta \EP}{\Delta \EP^{\mathrm{1d}}}
  = \exp\left[ -K\delta \left(
      \frac{1}{1+\e^{t_\mathrm{c}/(1+\Gamma)}}
      - \frac{1}{1+\e^{t_\mathrm{c}}}
    \right) \right].
\end{equation}
This is still a substantial benefit, although much less (for $\Gamma$
not too large) than that for rarely switching persisters, see
Eq.~\eqref{EPI-ratio-cat}.  Note that here $\Gamma$ only appears in the
effective catastrophe duration, not as the persister fraction.

Persisters are thus most valuable when stochastic switching is
relatively rare.  The fact that rare switching dominates in nature can
be attributed to an evolutionary process.

\section{Discussion and Conclusions}
\label{sec:discussion}

We have used a simple two-population model of normals and persisters to
show that (and how) persisters exponentially decrease the extinction
risk of an established bacterial population.  We have compared the
two-population system of normals and persisters to a normals-only system
starting from the same total population size.
Already in a constant environment favorable for normals, it is
beneficial to switch to the persister state: persisters contribute to
the MTE exponentially more than normals since their extinction is
delayed by first switching back.
When the population is under stress---that we model as a
catastrophe---the same buffering is effective, rendering persisters far
less prone to extinction, so that they exponentially reduce the EPI due
to the catastrophe.  For catastrophes which are long compared with the
lifetime of normals but short compared with the much longer switching
time scale (from persisters to normals), the reduction factor is squared
with respect to the MTE increase in a constant environment: persisters
are even more valuable for the population if it faces a catastrophe.

In exponential-growth models which focus on fitness, persisters are only
a burden in a constant favorable environment.  To explain their
existence with an overall benefit one needs to invoke temporary
adverse conditions.
In contrast, we have shown that persisters are always beneficial
as an insurance against the extinction of an established population, as
measured by the increased MTE, or by the reduced EPI during a
catastrophe, respectively.
We have also shown that to provide the optimal benefit, switching to and
from the persister state has to be rare compared with all other
processes.
In this sense, the extinction risk perspective presented here explains,
in a natural and robust way, the existence of persister phenotypes in
bacteria as well as the small switching rates from the normals to
persisters and back.

Our main analytical results~\eqref{MTE-ratio-nocat}
and~\eqref{EPI-ratio-cat} advocate that switching back from persisters
to normals should be rare compared with switching to the persister
state, leading to the largest possible fraction of persisters in the
metastable state (within the range where our theory applies).
For a bacterial population optimized solely against extinction
from the established state, this would be an intuitive strategy even in
favorable conditions.
During the growth stage, on the other hand, the population needs optimal
fitness to establish itself.
These two complementary strategies, optimizing two different quantities,
are not incompatible: the extinction risk perspective explains the mere
existence of persisters, already without invoking environmental
variations.
The switching rates themselves (fixing the metastable persister
fraction) may be tuned by evolution to optimize the growth stage in a
variable environment.

Our simple model neglects many features that can be biologically
relevant.
For example, in reality persisters have reduced but nonzero
birth and death rates.  Such a more realistic system still features
time-scale separation; persisters are now directly affected by a
catastrophe (e.g., inhibiting their births), but again on a much slower
scale than normals.  Therefore we expect a qualitatively similar
behavior.
Future work can attempt to account for the cost of switching to
persisters, for example, via competition between persisters and normals.
There are also many alternatives for the detailed dynamics during the
catastrophe.
For many of them, the WKB approximation to the master equation provides
a viable theoretical framework for determining the long-time behavior of
bacterial populations.

\begin{acknowledgments}
  We appreciate useful discussions with Nathalie Q.\ Balaban.  This work
  was supported by the Minerva foundation (IL), by the Israel Science
  Foundation (Grant No.~408/08), and by the U.S.-Israel Binational
  Science Foundation (Grant No.~2008075).
\end{acknowledgments}

%\bibliography{jrnlabrv,myrefs}
%Merlin.mbs v4.21 2009-07-09.
%

\end{document}